# Two-color atom guide and 1D optical lattice using evanescent fields of high-order transverse modes

Jian Fu, Xiang Yin, and Limin Tong

State Key Lab of Modern Optical Instrumentation, Department of Optical Engineering,
Zhejiang University, Hangzhou 310027, China

**Abstract**

We propose a two-color scheme of atom guide and 1D optical lattice using evanescent light fields of different transverse modes. The optical waveguide carries a red-detuned light and a blue-detuned light, with both modes far from resonance. The atom guide and 1D optical lattice potentials can be transformed to each other by using a Mach-Zehnder interferometer to accurately control mode transformation. This might provide a new approach to realize flexible transition between the guiding and trapping states of atoms.

## I. Introduction

The last few years have seen an increased interest in the design of neutral-atom nano traps and guides with optical planar waveguides and fibers. The atom traps and guides can be used as tools for atom interferometry [1], integrated atom optics [2], and quantum computation [3]. The idea of using evanescent waves (EW) to provide both attractive (red-detuning) and repulsive (blue-detuning) forces is due to Ovchinnikov *et al.* [4], who proposed using of two-color (i.e., red and blue detunings) evanescent waves and differing evanescent decay lengths to achieve an atom trap departing from a prism surface. Barnett *et al.* [5] proposed a two-color scheme to trap atoms using an EW above a single-mode, submicron optical channel waveguide. In the proposal, two polarizations are used to enlarger the differing evanescent decay lengths of the two-color lights. Two-color traps are sensitive to small field perturbations because the potentials defined by the red- and blue-detuned light are large, yet the net trapping potential is small. This sensitivity could be reduced by increasing the difference between the evanescent decay lengths of the two-color lights [6] or by using one-color scheme [7]. Christandl *et al.* [8] proposed an optical lattice scheme using the evanescent wave fields above an optical waveguide resulting from interference of different transverse modes. In the scheme, the one dimension (1D) optical lattice is constructed by destructively interfering waveguide modes of a channel or ridge waveguide and the 2D optical lattice is constructed by intersecting two of the 1D waveguide structures. It is necessary for producing the optical lattice potentials to accurately control different transverse modes excited in multimode waveguides. The accurate control could hardly be achieved by using a coupling grating [8].

In this paper, we discuss a two-color scheme of atom guide and 1D optical lattice using evanescent light fields of different transverse modes. The optical waveguide carries a red-detuned light and a blue-detuned light, with both modes far from resonance. By selectively exciting different transverse modes for the red-detuned light, we produce atom guide and 1D optical lattice potentials. Due to substantially differing evanescent decay length of different transverse modes, we produce a net trapping potential with a deep minimum, a large coherence time, and a large trap lifetime. By using the interference of the first-order and the fundamental modes, we produce 1D optical lattices along the optical waveguide. In order to accurately control the excitation of the two modes, we introduce a dual-mode waveguide Mach-Zehnder interferometer (MZI) to implement mode transformation between the two modes. The mode transformation provides a new approach to realize flexible transition between the guiding and trapping states of atoms, which would be desired in applications such as atom interferometer and quantum information.

This paper is organized as follows. In Sec. II, we discuss the two-color atom guide scheme using EW fields of high-order transverse mode above a submicrometer-width optical waveguide. In Sec. III, we discuss the 1D optical lattice using EW fields of mode superposition and the transformation between atom guide and optical lattice potentials. Our conclusions are given in Sec. IV.

**II. Two-color atom guide using EW fields of high-order transverse modes**

Ovchinnikov *et al.* [4] proposed using total internal reflection of blue- and red-detuned light off a planar surface to trap atoms. The difference in the decay lengths $L$ of the evanescent wave (EW) fields results in a potential minimum above the surface. In this section, we propose a two-color atom guide based on the EW fields of high-order transverse modes in a submicrometer-width waveguide. As shown in Fig. 1, our proposal is to utilize the different evanescent decay lengths of the fundament and the first-order modes carried in the dual-mode optical waveguide. Due to the larger difference in decay lengths of the two modes, we use far-off-resonance lights to produce a net potential with a deeper minimum, a larger coherence time, and a larger trap lifetime. In this section, we discuss in detail the atom guide potential that is defined by contributions from the optical dipole force of the EW fields, the surface van der Waals and Casimir interaction.

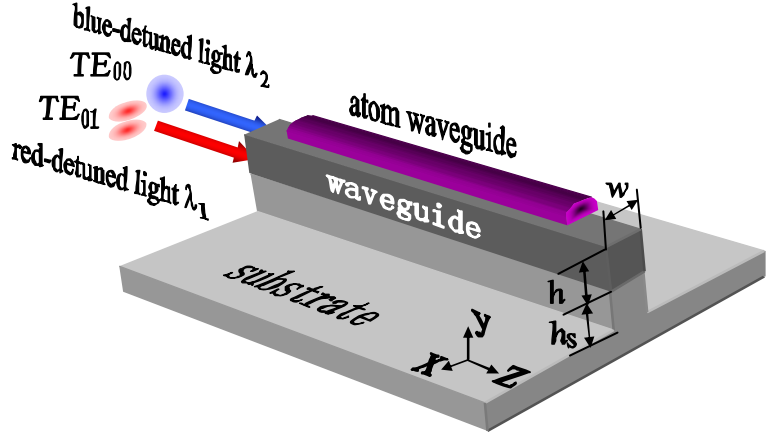

Fig. 1. Schematic diagram of a two-color atom waveguide.

### A. Theory of trapping potentials

The atoms in a near-resonant laser field of frequency $\omega$ can be exerted on a conservative force and a dissipative force. These forces, which are the results of momentum conservation in the absorption and reemission of photons, make different contributions to trapping atoms. The atom trapping potential is derived from the conservative force due to stimulated exchange of photons, whereas the dissipative force due to spontaneous photon emission is a mechanism for the loss of atoms from a trapping potential. Now we discuss in detail the two forces.

The conservative force is the gradient of a spatially dependent potential $U_{opt}(x,y,z)$ which can be expressed as the time-averaged induced dipole interaction energy in the electric field. The dipole potential $U_{opt}(x,y,z)$ for large detuning from resonance, $\Delta=\omega-\omega_0$ ($\Delta$ is greater than the excited state hyperfine splitting, but is much less than the atom transition frequency $\omega_0$), is given by

$$U_{opt}(x,y,z) = \frac{3\pi c^2}{2\omega_0^3}\frac{\Gamma}{\Delta}I(x,y,z), \qquad (1)$$

where the spontaneous decay rate of the excited state is $\Gamma = \left(\omega_0^3/3\pi\varepsilon_0\hbar c^3\right)\left|\langle e|\mu|g\rangle\right|^2$,

$\langle e|\mu|g \rangle$ is the dipole transition matrix element between ground $|g\rangle$ and excited $|e\rangle$ states, and $I(x,y,z)$ is the spatially dependent laser intensity. An important character for dipole trapping are immediately obvious from Eq. (1), that is the sign of the detuning $\Delta$ determines whether the potential is repulsive ($\Delta<0$, blue detuning) or attractive ($\Delta>0$, red detuning). Atoms in blue-detuned light are pushed towards the minimum of the field, whereas in red-detuned light the atoms are attracted towards the maximum of the field. Then we obtain the net dipole potential $U_{opt}(x,y,z)$ for the two-color light fields,

$$U_{opt}(x,y,z) = U_{opt}^{(b)}(x,y,z) + U_{opt}^{(r)}(x,y,z) = \frac{3\pi c^2 \Gamma}{2\omega_0^3}\left[\frac{I_b(x,y,z)}{|\Delta_2|} - \frac{I_r(x,y,z)}{|\Delta_1|}\right], \quad (2)$$

where $U_{opt}^{(b)}$, $U_{opt}^{(r)}$ are the dipole potentials and $I_b$, $I_r$ are the spatially dependent optical intensities for the blue- and red-detuned lights, respectively.

The dissipative force due to the absorption of mode photons and emission of other photons is a dissipative mechanism for trapping atoms. The dissipative atomic rate caused by a single light field is given by

$$\Gamma_{sc}(x,y,z) = \frac{3\pi c^2}{2\hbar\omega_0^3}\left(\frac{\Gamma}{\Delta}\right)^2 I(x,y,z). \quad (3)$$

From Eq. (1), we know that the rate has an additional factor $\Delta^{-1}$ as compared to the dipole potential $U_{opt}(x,y,z)$. But the rate for the two-color potential depends on the sum rather than the difference of the two field intensities, and has the form

$$\Gamma_{sc}(x,y,z) = \Gamma_{sc}^{(b)}(x,y,z) + \Gamma_{sc}^{(r)}(x,y,z) = \frac{3\pi c^2 \Gamma^2}{2\hbar\omega_0^3}\left[\frac{I_b(x,y,z)}{\Delta_2^2} + \frac{I_r(x,y,z)}{\Delta_1^2}\right], \quad (4)$$

where $\Gamma_{sc}^{(b)}$, $\Gamma_{sc}^{(r)}$ are the spontaneous scattering rates for the red- and blue-detuned light, respectively. The dissipative atomic rate limits the coherence time of the trap. For atoms close to the trap minimum position $x_m, y_m, z_m$ of the two-color potential $U_{opt}$, the characteristic coherence time is

$$\tau_{coh} = \frac{1}{\Gamma_{sc}(x_m, y_m, z_m)}. \quad (5)$$

Every scattered photon imparts a recoil energy $E_r = (\hbar\omega/c)^2/2m$ to the atom, where $m$ is the mass of the atom. Due to the photon recoil heating, the trap lifetime of atoms trapped in the potential $U_D \equiv U_{opt}(x_m, y_m, z_m)$ can be characterized by

$$\tau_{trap} = \frac{U_D}{2\left[E_r^{(r)}\Gamma_{sc}^{(r)}(x_m, y_m, z_m) + E_r^{(b)}\Gamma_{sc}^{(b)}(x_m, y_m, z_m)\right]}, \quad (6)$$

where $E_r^{(b)}$, $E_r^{(r)}$ are the recoil energies for the blue- and red-detuned lights, respectively. Due to the additional factor $\Delta^{-1}$, the dissipative rates can be effectively reduced for large detuning. Therefore, dipole traps usually work with light far-off resonance and high intensities to maintain a certain trap depth while reducing the dissipative rate to a negligible value.

Atoms near a flat dielectric surface undergo a position dependent energy shift that gives rise to a force that attracts the atom towards the surface. It is known as the van der Waals force ($l \ll \lambda/2\pi$) or the Casimir force ($l \gg \lambda/2\pi$), depending on the distance $l$ from the surface compared to the transition wavelength $\lambda$ of the atoms. There is a smooth cross-over from van der Waals ($U \sim l^{-3}$) to Casimir ($U \sim l^{-4}$). We use an interpolation formula, which bridges the two regimes and retains the respective power law behavior in each limit, to characterize the atom-surface potential [9],

$$U_{sur} = \left[(1+1.098z)^{-1} - 0.00493z\left(1+0.00987z^3 - 0.00064z^4\right)^{-1}\right]\frac{\varepsilon-1}{\varepsilon+1}\frac{C^{(3)}}{z^3}, \quad (7)$$

where $z = 2\pi l/\lambda$ is the rescaled distance between atoms and the surface, $\varepsilon$ is the dielectric permittivity, and $C^{(3)} = 0.113\hbar\Gamma/\lambda^3$ is related to the transition dipoles for $^{87}$Rb atoms. In order to keep the atoms from sticking to the waveguide surface, it is necessary to use the blue-detuned light to stand against the surface interaction.

Other contributions, such as the gravitational potential, are completely negligible over the distances that will be of interest ($mg=0.1\mu K/\mu m$ for $^{87}$Rb atoms). The total trapping potential $U$ is the sum of the net optical potential $U_{opt}$ and the atom-surface potential $U_{sur}$.

### B. High-order transverse mode and its evanescent field

To produce a net trapping potential with a large depth, as well as a large coherence time and a large trap lifetime, we need sufficiently differing the decay lengths of the red- and blue-detuned light fields. In the previous schemes [5-7], the difference is caused by different wavelengths or polarization modes of light fields carried in a single-mode optical waveguide. In this paper, our proposal is differing evanescent decay lengths of two different transverse modes carried in the dual-mode optical waveguide. In the following text we discuss in detail the fundamental mode and the first high-order mode and their evanescent fields.

Considering a channel waveguide as shown in Fig. 1 that has a core rib of width $w$, height $h$ and refractive index (RI) $n_g$ deposited on a substrate step of height $h_s$ and RI $n_s$ with an infinite vacuum clad of refractive index 1, an optical field in the propagation direction, longitudinal $z$ direction, is restricted within the core region. Generally, the optical field distribution is classified under two types: transverse-magnetic (TM) mode and transverse-electric (TE) mode. Considering a linearly polarized TE mode, we have by definition $E_z = 0$ and the electric-field component of the envelope $\phi(x,y,z) = E(x,y)e^{i\beta z}$, that can be obtained as a solution of the reduced wave equation

$$\left(\nabla_x^2 + \nabla_y^2\right)E + \left[n^2(x,y)k_0^2 - \beta^2\right]E = 0, \tag{8}$$

where $\beta$ is the propagation constant, $n(x,y)$ is the spatially dependent refractive index, and $k_0 = 2\pi/\lambda$ is the free space wave number. Solving the equation directly subject to the boundary conditions of the waveguide structure, the electric-field component $\phi(x,y,z)$ can be expressed as the superposition of guide modes $E_n(x,y)$ corresponding to the discrete propagation constants $\beta_n$,

$$\phi(x,y,z) = \sum_n C_n E_n(x,y)e^{i\beta_n z}, \tag{9}$$

where $C_n$ are the mode superposition coefficient, and the optical intensity is related to the squared amplitude of the electric field component, $I(x,y,z) = \tfrac{1}{2}\varepsilon_0 c |\phi(x,y,z)|^2$.

In our scheme, a dual-mode optical waveguide is employed to carry the red- and blue-detuned light fields. The width $w$ and height $h$ of the optical waveguide are carefully selected to guarantee that there are exactly two guide modes: the fundamental mode $TE_{00}$ (the subscripts denote the mode numbers of $x$ and $y$, respectively) and the first high-order mode. In the square waveguide, there are two orthogonal and degenerate first high-order modes, namely $TE_{01}$ and $TE_{10}$ modes. In order to produce trapping potentials on the top of waveguides, the $TE_{01}$ mode is selected in our scheme. The discussion of the cutoff condition for guide modes is presented in many places [10], here we investigate a waveguide fabricated in silicon on isolator (SOI) utilizing the results. The SOI waveguide with a guide index $n_g$=3.42 and a substrate index $n_s$=1.45, is inherently multimode because of the high refractive index difference between Si and SiO$_2$. According to the cutoff condition, we appropriately tailor its width and height, for example $w=h=0.3\mu$m, which are large enough to carry the lowest two modes (the TE$_{00}$ and TE$_{01}$ modes in our case) for the red-detuned light (865nm), yet small enough to cutoff higher order modes than the TE$_{01}$ mode.

It is noticeable that the evanescent fields for the transverse modes provide a large difference of evanescent decay lengths. Using numerical simulations, we obtain the normalized intensity distributions in *x-y* plane (as shown in Fig. 2), and the propagation constants $\beta_0 = 22.04\times10^6 m^{-1}$ and $\beta_1 = 17.23\times10^6 m^{-1}$, respectively for the TE$_{00}$ and TE$_{01}$ modes. The maximal value of the intensity distribution of the TE$_{00}$ mode is at the center of the waveguide, but the maximal values of the TE$_{01}$ mode are near the edges of the waveguide. This leads to substantially different evanescent waves of the two modes. In order to quantify the difference, we calculate the evanescent decay lengths for the TE$_{00}$, TM$_{00}$ and TE$_{01}$ modes as a function of light wavelength $\lambda$, as shown in Fig. 3. Apparently, a larger decay-length difference can be obtained than that for the case of different polarizations carried in the single-mode optical waveguide [5]. In general, a larger difference in decay lengths will provide a stronger trap for a given laser power as well as a smaller spontaneous decay rate at a given trap depth and detuning.

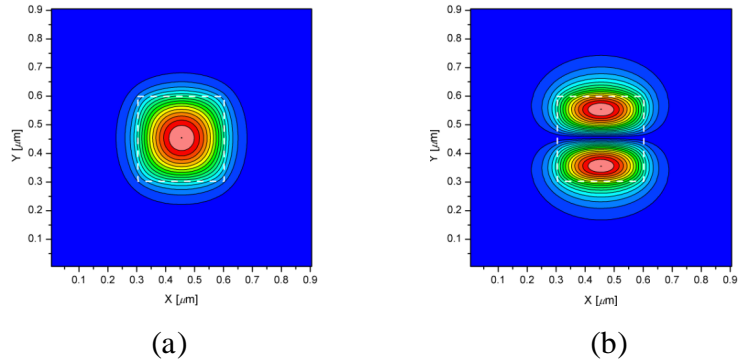

(a)            (b)

Fig. 2. The normalized intensity distributions in *x-y* plane for (a) the TE$_{00}$ mode and (b) the TE$_{01}$ mode of the light $\lambda$ = 865nm, where the dash lines denote the geometry of the core rib ($w=h=0.3\mu$m) as shown in Fig. 1.

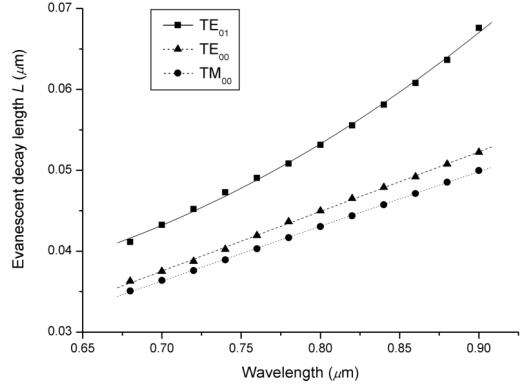

Fig. 3. The evanescent decay lengths for the $TE_{00}$, $TM_{00}$ and $TE_{01}$ modes as a function of light wavelength, where $L$ are the characteristic lengths for the exponential decay electric fields $E \propto e^{-y/L}$.

**C. Atom guide potentials**

In this work, we consider trapping ground-state $^{87}$Rb atoms in their ground-state manifold $S_{1/2}$. Neglecting hyperfine structure, the atoms have two strong transitions, at 795 nm ($P_{1/2}$ line) and 780 nm ($P_{3/2}$ line). To trap the atoms, we use red- and blue-detuned lights with wavelengths $\lambda_1 = 865$ nm and $\lambda_2 = 700$ nm, respectively. To consider the fine-structure energy splitting, we substitute $\Gamma_{1/2}/3\Delta_{1/2} + 2\Gamma_{3/2}/3\Delta_{3/2}$ for $\Gamma/\Delta$ in Eq. (1) and Eq. (2), where the spontaneous decay rates $\Gamma_{1/2}, \Gamma_{3/2}$ for $P_{1/2}$ and $P_{3/2}$ are 36.1 MHz and 38.1MHz, respectively. In order to avoid the effect of substrate attractive potentials, the $TE_{01}$ mode is excited for the red-detuned light to produce a trapping potential on the top of optical waveguide. The total trapping potential $U$ of the atoms is the sum of the net optical potential $U_{opt}$ and the atom-surface potential $U_{sur}$, i.e.,

$$U(x,y,z) = U_{opt} + U_{sur}, \qquad (10)$$

We use Eqs. (2) and (7) to calculate the trapping potential $U$ of the ground-state $^{87}$Rb atoms outside the rib waveguide with the red- and blue-detuned light fields.

In order to achieve a larger decay-length difference, the $TE_{00}$ mode is selectively excited for the blue-detuned light, and the $TE_{01}$ mode is selectively excited for the red-detuned light. By using numerical simulation, we obtain that the corresponding evanescent decay lengths $L_R = 0.0617\mu m$ (the red-detuned light, $TE_{01}$ mode) and $L_B = 0.0375\mu m$ (the blue-detuned light, $TE_{00}$ mode). The relative difference between the decay lengths is measured by the parameter $\alpha_L = (L_R - L_B)/L_B \approx 0.65$. The obtained value of this parameter is larger than the characteristic value $\alpha_L \approx 0.36$ estimated for the case of different polarizations [5], such as $TE_{00}$ and $TM_{00}$ modes.

In Fig. 4, we present an example of a two-dimensional trapping potential in (a) $x$-$y$ plane and (b) $y$-$z$ plane outside the optical waveguide. Assuming an input of 1.5 $mW$ of the red-detuned light ($TE_{01}$, 865 nm) and 40 $mW$ of the blue-detuned light ($TE_{00}$, 700 nm), we find a potential minimum $x_{min} \approx 0.092 \mu m$ departed from the waveguide surface with trapping depths of $U_D \approx 114.6 \mu K$. In the trapping potential, atoms are confined in $x$ and $y$ dimensions and allowed free propagation in $z$ direction. We estimate some critical trapping parameters for the case of Fig. 4. The rates of scattering due to the trapping fields at the potential minimum are $\Gamma_{sc}^{(r)} \approx 6.1 s^{-1}$ and $\Gamma_{sc}^{(b)} \approx 2.7 s^{-1}$ for the red- and blue-detuned light fields, respectively. According to Eq. (3), the light scattering rate $\Gamma_{sc} \approx 8.8 s^{-1}$ and the coherence time $\tau_{coh} \approx 113.6 ms$ can be obtained. Due to the recoil heating $E_r^{(r)} = 0.145 \mu K$, $E_r^{(b)} = 0.221 \mu K$ for the red- and blue-detuned lights respectively, we characterize the trap lifetime $\tau_{trap} \approx 77.6 s$ for the trap depth $U_D \approx 114.6 \mu K$. At the trapping minimum, the transverse oscillation frequencies are $\omega_x / 2\pi \approx 51 kHz$, $\omega_y / 2\pi \approx 299 kHz$, giving an estimate of about 14.3 $\mu K$ for the energy of the atomic mode spacing. These frequencies are much larger than previously demonstrated with magnetic wire guides [11] and provide a ground-state localization of roughly $l_x = 39 nm$ and $l_y = 16 nm$.

Given the optical waveguide, we are free to choose three experimental parameters, namely the optical powers carried in the two modes, and the detuning Δ. The optical powers can be expressed as total power $P_{tot} = P_{red} + P_{blue}$, and the power ratio $R = P_{red} / P_{blue}$. The potential shape will be affected by $R$ alone: an increase in $R$ leads to an increase in the depth $U_D$ of the trapping potential and to a shift of the local minimum point $x_{min}$ towards the fiber surface, otherwise, an decrease in $R$ leads to a decrease in the depth $U_D$ of the trapping potential and to a shift of the local minimum point $x_{min}$ farther away from the fiber surface. Therefore, to produce a potential with a deep trapping minimum outside the fiber and with a high repulsive wall in the region of $x < x_{min}$, the ratio $R$ must be optimized appropriately. We increase the depth and harmonic frequencies of the trapping potential at the expense of increased atomic light scattering. The position and value of trap minimum, atomic light scattering rates, and the trap lifetimes determined by atom recoil heating are given in Table I for several different power ratios $R$. It is noticeable that the power ratios are quite small for the power of the red-detuned light is much smaller than the blue-detuned light. This is due to the large difference between their evanescent decay lengths.

Compared to the schemes utilizing different polarizations, the two-color light fields with different transverse modes provide a net trapping potential with a deeper minimum, a larger coherence time, and a larger trap lifetime.

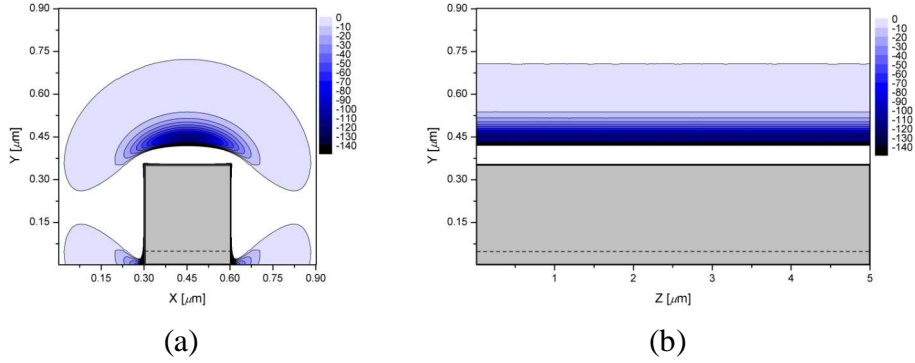

(a)            (b)

Fig. 4. A two-color trapping potential of atom guide in (a) *x-y* plane and (b) *y-z* plane. The waveguide structure are shown in Fig. 1. A maximum trapping depth of 114.6 $\mu K$ is achieved using 1.5 $mW$ of red light (TE$_{01}$, 865 $nm$) and 40 $mW$ blue light (TE$_{00}$, 700 $nm$).

TABLE I. Variation of trapping potentials with red power $P_{red}$ for the example shown in Fig. 4: vertical position of trap minimum $x_{min}$, maximum trapping depth $U_D$, atomic light scattering rate $\Gamma_{sc}$ at $x_{min}$, and trap lifetime $\tau_{trap} = U_D \big/ \left( E_r^{(r)} \Gamma_{sc}^{(r)} + E_r^{(b)} \Gamma_{sc}^{(b)} \right)$. The input power of blue-detuned light $P_{blue}$ is fixed at 40 $mW$.

| $P_{red}$ ($mW$) | $x_{min}$ ($\mu m$) | $U_D$ ($\mu K$) | $\Gamma_{sc}$ ($s^{-1}$) | $\tau_{trap}$ ($s$) |
|---|---|---|---|---|
| 0.5 | 0.137 | 7.49 | 0.56 | 12.4 |
| 1.0 | 0.107 | 41.51 | 3.14 | 78.6 |
| 1.5 | 0.092 | 114.6 | 8.78 | 77.6 |
| 2.0 | 0.075 | 236.9 | 18.5 | 74.7 |
| 2.5 | 0.066 | 417.4 | 31.9 | 77.7 |

### III. 1D lattice using evanescent light fields of mode superposition

In this section, we discuss a scheme to produce a 1D lattice utilizing evanescent fields of the red-detuned light with a superposition of the TE$_{00}$ and TE$_{01}$ modes. Different from the scheme of standing waves, the counter propagating arrangement is not required in our proposal. We discuss arranging the red- and blue-detuned light field to produce the 1D lattice and the transformation between the 1D lattices and the atom guides.

### A. Mode superposition and its periodic intensity distribution

From the model field analysis, we know that the fundamental mode $TE_{00}$ is even mode and the first-order mode $TE_{01}$ is odd mode. According to Eq. (2), we express arbitrary TE optical fields carried in the dual-mode waveguide as a superposition of the $TE_{00}$ and $TE_{01}$ modes,

$$\phi(x,y,z) = C_0 E_0(x,y) e^{i(\beta_0 z + \theta_0)} + C_1 E_1(x,y) e^{i(\beta_1 z + \theta_1)}, \tag{11}$$

where $E_0(x,y)$ and $E_1(x,y)$ are the electric-field components, $\theta_0$ and $\theta_1$ are the phases of $TE_{00}$ and $TE_{01}$ modes, respectively. Then, the optical intensity distribution of the mode superposition can be obtained,

$$\begin{aligned} I(x,y,z) &= \tfrac{1}{2}\varepsilon_0 c \phi(x,y,z) \phi^*(x,y,z) \\ &= \tfrac{1}{2}\varepsilon_0 c [|C_0|^2 |E_0(x,y)|^2 + |C_1|^2 |E_1(x,y)|^2 \\ &\quad + C_0 C_1^* E_0(x,y) E_1^*(x,y) e^{i(\Delta\beta z + \Delta\theta)} + C_0^* C_1 E_0^*(x,y) E_1(x,y) e^{-i(\Delta\beta z + \Delta\theta)}], \end{aligned} \tag{12}$$

with the propagation constant difference $\Delta\beta = \beta_0 - \beta_1$ and the relative phase $\Delta\theta = \theta_0 - \theta_1$. Apparently the last two terms in Eq. (12) result from the coherent superposition of the two modes, which contain a periodic function of $z$ (the propagation direction). This leads to the intensity distribution periodically varying (a fringe pattern) along the propagation direction. The fringe pattern can be moved by modulating the relative phase $\Delta\theta$. Therefore, the intensity distribution may provide periodical 3D atomic confinements.

In order to demonstrate the property of mode superposition, we numerically simulate the propagation of the red-detuned light in superposition of the $TE_{00}$ and $TE_{01}$ modes by using the finite differential beam propagation method (FD-BPM) [12]. The result shown in Fig. 5 demonstrates a typical variation of the intensity of the mode superposition along the propagation direction $z$. Apparently the intensity distribution varies periodically along the propagation direction $z$ with a period of $2\pi/\Delta\beta = 1.31\mu m$. Further, we numerically calculate the normalized intensity distributions of $x$-$y$ cross section for the mode superposition at three different positions along $z$ in a half period with the results shown in Fig. 6. Different from the cases of the single modes, the evanescent waves of the mode superposition also periodically varies along the propagation direction $z$. We utilize the periodic optical intensity to produce a periodic potential to trap and guide atoms, which forms a 1D optical lattice along the optical waveguide.

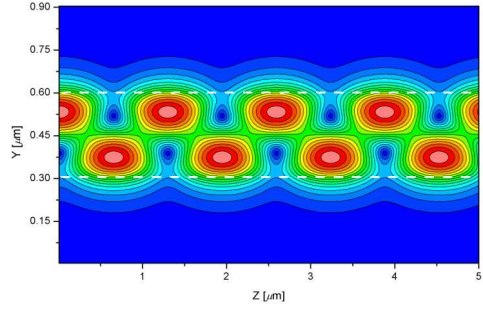

Fig. 5. A typical intensity variation of the red-detuned light in mode superposition along the propagation direction $z$ with a variation period of 1.31 $\mu$m. The dashed lines denote the core geometry of the waveguide given in Fig. 1.

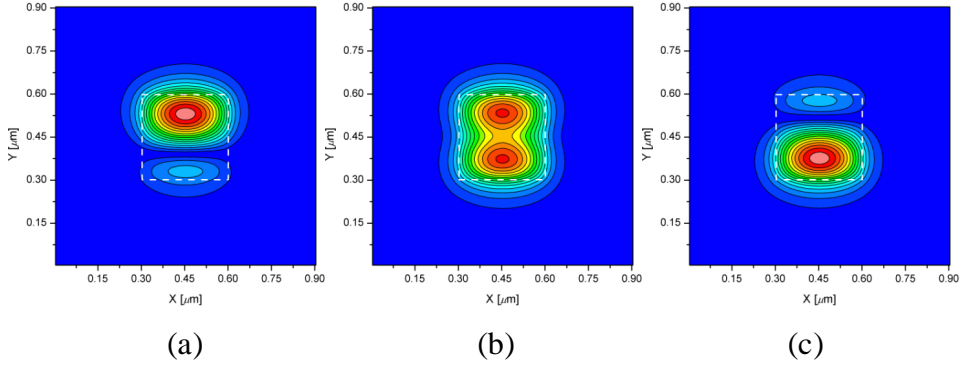

(a)           (b)           (c)

Fig. 6. The normalized intensity distributions in $x$-$y$ plane for the mode superposition at (a) $z = 0\ \mu$m, (b) $z = 0.325\ \mu$m, (c) $z = 0.65\ \mu$m.

**B. 1D optical lattice potential**

In order to produce a 1D lattice, we utilize the red-detuned light excited in the superposition of the $TE_{00}$ and $TE_{01}$ modes and the blue-detuned light excited in the $TE_{00}$ mode. The intensity distribution of red-detuned light introduces an attractive potential periodically varying along the propagation direction $z$ that forms a 1D lattice.

Assuming the input powers and wavelengths of the red- and blue-detuned lights to be the same as the last section, we present an example of the 1D lattice in Fig. 7. A series of periodical potential minimums along $z$ can be found at $x_{min} \approx 0.08 \mu m$ departed from the waveguide surface with trapping depths of $U_D \approx 146 \mu K$ and the variation period is same as the variation period $1.31 \mu m$ of the intensity distribution. Further, we obtain the trapping potential distributions in $x$-$y$ plane at three different positions along $z$ with the results shown in Fig. 8. Apparently, the trapping potential provides three dimensional atom confinement. We also estimate some critical trapping parameters for the trapping potential. The rates of scattering due to the trapping fields at the potential minimum are $\Gamma_{sc}^{(r)} \approx 9.19 s^{-1}$ and $\Gamma_{sc}^{(b)} \approx 4.53 s^{-1}$. According to Eq. (3), the light scattering rate $\Gamma_{sc} \approx 13.71 s^{-1}$ and the coherence time $\tau_{coh} \approx 73 ms$ can be obtained. We characterize the trap lifetime $\tau_{trap} \approx 118.7 s$ for the trap depth $U_D \approx 146 \mu K$. At the trapping minimum, the transverse oscillation frequencies are $\omega_x / 2\pi \approx 56 kHz$, $\omega_y / 2\pi \approx 346 kHz$ and $\omega_z / 2\pi \approx 32 kHz$, giving an estimate of about 16.6 $\mu K$ for the energy of the atomic mode spacing and providing for a ground-state localization of roughly $l_x \approx 36 nm$, $l_y \approx 15 nm$ and $l_z \approx 40 nm$.

In conclusion, we have shown that it is possible to trap atoms in three dimension and to form the 1D optical lattice utilizing the evanescent fields of the red-detuned light excited in the superposition of the $TE_{00}$ and $TE_{01}$ modes and the blue-detuned light excited in the $TE_{00}$ mode. And the trapping potential might be more robust than other two-color potentials against the coherent reflections of lights.

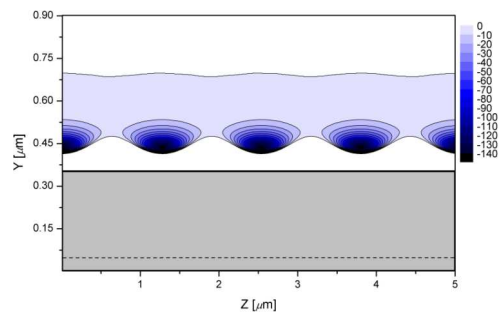

Fig. 7. A trapping potential of the 1D optical lattice periodically varying along the direction $z$ is shown.

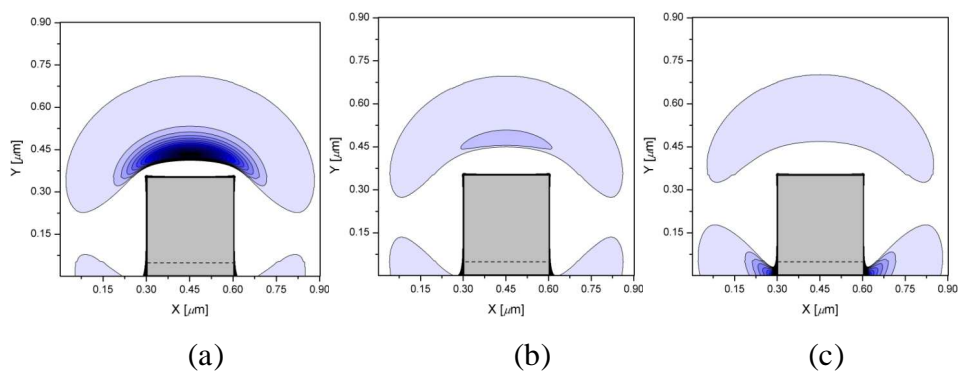

(a)                      (b)                      (c)

Fig. 8. The trapping potential distributions in $x$-$y$ plane at (a) $z = 0\mu m$, (b) $z = 0.325\mu m$, (c) $z = 0.65\mu m$, is shown. Contours are the same as in Fig. 11.

## C. Transformation between the atom guide and the 1D optical Lattice

In order to realize the transformation between the atom guide and the 1D optical Lattice, we need some approaches to accurately control the modes of light fields carried in waveguides. Here, we propose a dual-mode waveguide MZI to implement mode transformation between the $TE_{00}$ and $TE_{10}$ modes. The MZI contains three stages: an initial Y-splitter which splits the input fields, a central section where one of the waveguides contains a phase modulator, and another Y-combiner which recombines the signals at the output waveguide. It is noticeable that the $TE_{00}$ mode is symmetric and the $TE_{10}$ mode is anti-symmetric. When a $TE_{00}$ mode light field is launched into the MZI, the light field is split into two symmetric parts by the Y-splitter. Then a $\pi$-phase difference between the two parts is induced by using the phase modulator. The two anti-symmetric parts are finally recombined by the Y-combiner and the output field become the $TE_{10}$ mode. If the mode disturbances and intensity losses induced by the MZI are neglected, the mode transformation can be described as an unitary transformation,

$$U = \begin{pmatrix} \cos(\theta/2) & i\sin(\theta/2) \\ i\sin(\theta/2) & \cos(\theta/2) \end{pmatrix}, \tag{13}$$

where the phase $\theta$ accounts for any phase difference between the two arms of the MZI. In this matrix picture, the $TE_{00}$ mode is denoted by a unit vector $(1\ 0)^T$ and the $TE_{10}$ mode denoted by $(0\ 1)^T$. Assuming any mode superposition $\phi_{in} = (C_0^i\ C_1^i)^T$ as the input field of the MZI, we obtain the output field as follow,

$$\begin{aligned}\phi_{out} &= U\phi_{in} \\ &= \begin{pmatrix} \cos(\theta/2)C_0^i + i\sin(\theta/2)C_1^i \\ \cos(\theta/2)C_1^i + i\sin(\theta/2)C_0^i \end{pmatrix}.\end{aligned} \tag{14}$$

When the field in the $TE_{00}$ mode $\phi_{in} = (1\ 0)^T$ is inputted, we obtain the $TE_{10}$ mode by setting the phase difference $\theta=\pi$, and obtain the mode superposition by setting $\theta=\pi/2$. In order to validate the theoretic analysis, we perform detailed numerical simulations of the MZI by using FD-BPM. The MZI is composed of the dual-mode waveguides mentioned in the last section and the wavelength of the launched light field is 1.06 $\mu$m. In the simulation, the phase differences $\theta$ can be obtained by properly modulating the core refractive index. When the length of the phase modulator $l = 50\ \mu m$ and the RI modulating difference $\Delta n = 0.01012$, the phase difference $\theta = \pi$ and the $TE_{00}$ mode is transformed to the $TE_{10}$ mode, as shown in Fig. 9 (a). When $\Delta n = 0.00506$, the phase difference $\theta = \pi/2$ and the

TE$_{00}$ mode is transformed to the mode superposition, as shown in Fig. 9 (b). Further, we obtain the relation of the mode superposition coefficients and the refractive index difference, as shown in Fig. 10. Apparently, any superposition of the TE$_{00}$ and TE$_{10}$ modes can be obtained by properly modulating the phase difference. It provides a rapid and accurate approach to transiting between the atom guide and the 1D lattice. Figure 11 presents a middle state of the atom guide and the 1D lattice with the phase difference $\theta = 5\pi/6$.

It is noticeable that the TE$_{01}$ mode, rather than the TE$_{10}$ mode, is selectively excited for the red-detuned light in our scheme. The two modes are orthogonal and degenerate (have the same propagation constant) in the square waveguide and their field distributions are twisted by 90°. There are many methods to realize the mode conversion of the TE$_{10}$ and TE$_{01}$ modes, e.g. using a twisted waveguide [13]. Similar to the twisted waveguide schemes, we present here a simple scheme to convert the modes by using a directional coupler composed of two parallel off-plane waveguides, as shown in Fig. 12.

In conclusion, we have shown an approach to realize the transformation between an atom guide and 1D optical lattice by accurately controlling the modes of light fields carried in waveguides. Apparently, this approach has a distinct advantage over the coupling grating scheme proposed in Ref. [8].

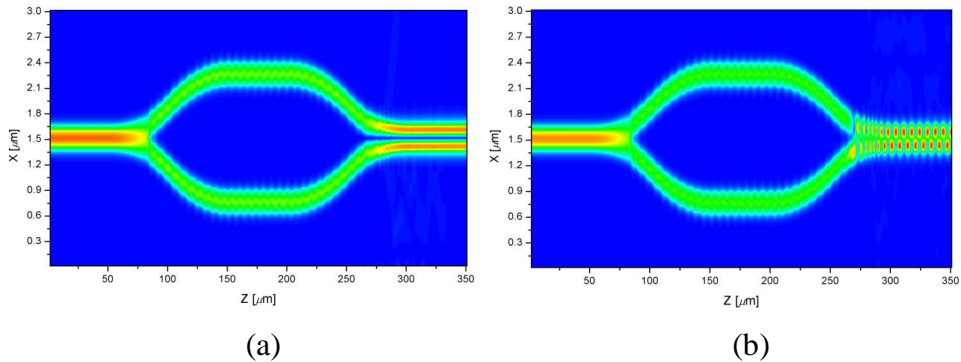

(a)        (b)

Fig. 9. FD-BPM simulations for the mode transformations from TE$_{00}$ mode to (a) TE$_{10}$ mode when $\Delta n = 0.01012$, and (b) the mode superposition when $\Delta n = 0.00506$ with the modulator length $l = 50 \mu m$.

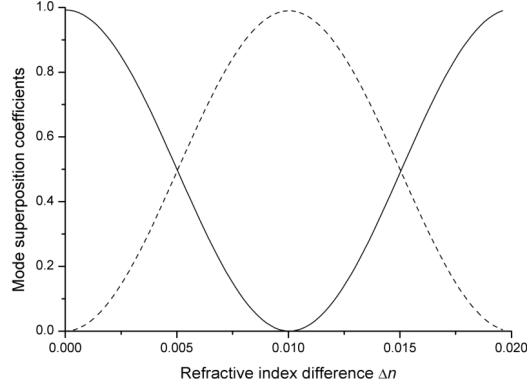

Fig. 10. Mode superposition coefficients of the $TE_{00}$ mode (the solid line) and the $TE_{10}$ mode (the dashed line) as functions of the RI modulating difference $\Delta n$.

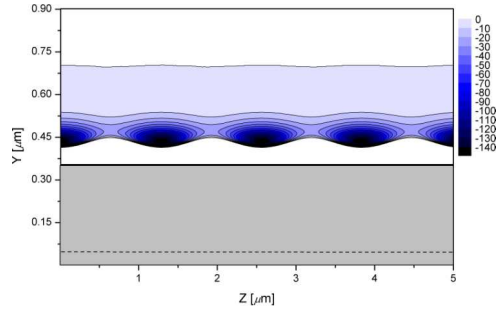

Fig. 11. A middle state of the atom guide and the 1D lattice. The input powers and wavelengths of the red- and blue-detuned lights are the same as used in Fig. 7, and the phase difference $\theta$ is $5\pi/6$.

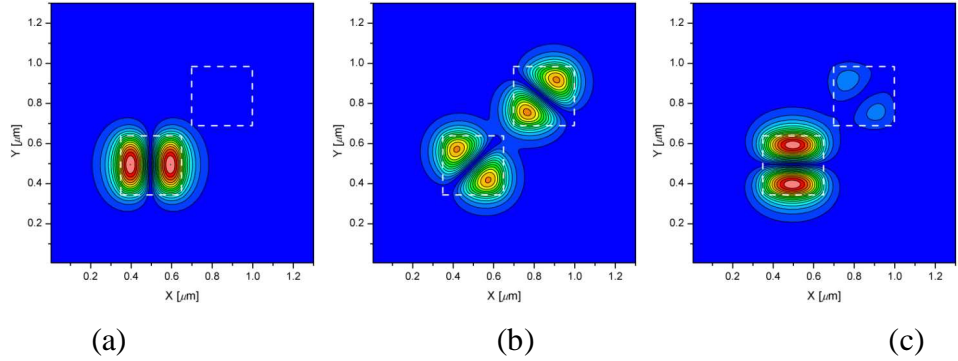

(a)                      (b)                      (c)

Fig. 12. The $TE_{10}$-$TE_{01}$ mode conversion using a directional coupler. The directional coupler is composed of two parallel off-plane waveguides (denoted by the dashed lines) with the waveguide gap $w_g$=0.042 $\mu$m and the couple length $L_c$=24.38 $\mu$m. The figures demonstrate the intensity distributions in $x$-$y$ plane for the mode conversion process at (a) $z$=0, (b) $z$=12.19 $\mu$m, (c) $z$=24.38 $\mu$m.

## IV. Conclusions

In this paper, we have discussed two-color atom guide and 1D optical lattice schemes using evanescent light fields of high-order transverse modes above a submicron waveguide. Due to a larger difference in decay lengths, we produce a stronger trap with a given laser power, as well as a smaller spontaneous decay rate at a given trap depth and detuning. In order to accurately control the excitation of different modes, we propose mode transformation using a dual-mode waveguide MZI. The mode transformation provide a new approach to realize flexible transition between the guiding and trapping states of atoms. It is interesting to mention that large decay-length differences provided by different transverse modes may greatly increase the robustness of the two-color trap to light scattered by defects, junctions, etc. in the waveguide. This will be the subject of further investigation.

## Acknowledgement
This work is supported by National Natural Science Foundation of China under Grant No. 60407003. The authors would like to thank Dr. Qiang Lin and Mr. Xiaoshun Jiang for valuable discussions.